%% file: main.tex
\documentclass[10pt,a4paper,conference]{IEEEtran}
\ifCLASSINFOpdf
  \usepackage[pdftex]{graphicx}
  \usepackage[bookmarks=false]{hyperref}
  \usepackage{caption}
  \usepackage{subcaption}
  \usepackage{color}
\else
  \usepackage[dvips]{graphicx}
\fi
\usepackage[usenames,dvipsnames,svgnames]{xcolor}

\newcommand{\optional}[1]{\ignorespaces}

\begin{document}
%
\title{D-STREAMON - a NFV-capable distributed framework for network monitoring}

\author{\IEEEauthorblockN{
Pier Luigi Ventre\IEEEauthorrefmark{1},
Alberto Caponi\IEEEauthorrefmark{2},
Davide Palmisano\IEEEauthorrefmark{1},
Stefano Salsano\IEEEauthorrefmark{1}\IEEEauthorrefmark{2},
Giuseppe Siracusano\IEEEauthorrefmark{1},\\
Marco Bonola\IEEEauthorrefmark{1},
Giuseppe Bianchi\IEEEauthorrefmark{1}\IEEEauthorrefmark{2}
}
\IEEEauthorblockA{
\IEEEauthorrefmark{1}University of Rome Tor Vergata, Italy,
\IEEEauthorrefmark{2}CNIT, Italy,
}
Email: \{name.surname\}@uniroma2.it
\\
\\
Submitted paper - June 2016
}


%


\maketitle

\input{inc/abstract}

%
\IEEEpeerreviewmaketitle

\vspace{1.5ex}
\vspace{-2ex}

\input{inc/introduction}

\input{inc/streamon}

\input{inc/dstreamon}

\input{inc/experimentalresults}

\input{inc/relatedwork}

\input{inc/conclusions}


\section*{Acknowledgment}
This research was partially supported by the EU Commission within the Horizon 2020 program, SCISSOR project grant no 644425.



%

\bibliographystyle{IEEEtran}
\bibliography{main}

\end{document}

%% file: inc/abstract.tex
\begin{abstract}
Many reasons make NFV an attractive paradigm for IT security: lowers costs, agile operations and better isolation as well as fast security updates, improved incident responses and better level of automation. At the same time, the network threats tend to be increasingly complex and distributed, implying huge traffic scale to be monitored and increasingly strict mitigation delay requirements. Considering the current trend of the networking and the requirements to counteract to the evolution of cyber-threats, it is expected that also network monitoring will move towards NFV based solutions. In this paper, we present Distributed StreaMon (D-StreaMon) an NFV-capable distributed framework for network monitoring. D-StreaMon has been designed to face the above described challenges. It relies on the StreaMon platform, a solution for network monitoring originally designed for traditional middleboxes. An evolution path which migrates StreaMon from middleboxes to Virtual Network Functions (VNFs) is described. The paper reports a performance evaluation of the realized NFV based solution and discusses potential benefits in monitoring tenants' VMs for Service Providers.
\end{abstract}


%% file: inc/introduction.tex
\section{Introduction}
The fast evolving complexity and heterogeneous nature of modern cyber-threats and network monitoring as well as the increasing interest in virtualization approaches for more complex network middlebox functionalities call for new scalable, accurate and flexible solutions to virtualize and simplify the programming and deployment of online (i.e. stream-based) traffic analysis functions. The real challenge is to promptly react to the mutating needs by deploying custom traffic analyses functionalities, capable of tracking events and detect different behaviours of attacks. Such objectives can be reached by efficiently handle the many heterogeneous features, events, and conditions which characterize an operational failure, a network's application mis-behavior, an anomaly as well as an incoming attack. Such needed level of flexibility and programmability should address scalability by design, through systematic exploitation of stream-based analysis techniques. And, even more challenging, traffic analyses and mitigation primitives should be ideally brought inside the monitored network and the monitoring probes themselves. This permits to avoid the centralization of the analysis that requires exporting traffic data to a central point that results to be an inadequate way to cope with the huge traffic scale and the strict (ideally real-time) mitigation delay requirements. 

The above challenges fit well in the current trends towards the softwarization of networks~\cite{galis}~\cite{kind} which includes technologies like Network Function Virtualization (NFV)~\cite{nfv} and Software Defined Networking (SDN)~\cite{sdn}. More specifically, NFV, focusing on the decoupling of network functions from physical devices on which they run, becomes the ideal technology to bring to the reality the idea of "spreading" network monitoring inside the infrastructure. NFV leverages on virtualization technologies to execute these network appliances in virtual resources deployed in commodity servers. Its promises are the achievement of unmatched infrastructure scalability, flexibility and efficiency of networks (reducing equipment and operational costs) through the ubiquitous employment of software-based network appliances on COTS hardware. 
As all new emerging technologies, NFV introduces new security issues and challenges to be addressed. Since the software is more fragile with respect to the hardware, it offers to an attacker more features to exploit. It implies that the shift of more functions to the software even increase the chance of disasters. On the other hand, the software enables to streamline the procedure of security updates and in general the impact of deploying security updates.
In this direction, NFV delivers agile security operations and better isolation: virtual resources are logically separated, if necessary some Virtual Network Functions (VNFs) can be deployed on isolated nodes or in nodes that match some security-pertinent criteria.
Unmatched incident response thanks to the inherent flexibility and better level of automation which can guarantee among other things rapid and flexible re-configuration of virtual network appliances.
We expect that network security solutions could benefit in moving towards NFV paradigm, where security systems or middleboxes are substituted by software components running in commodity server. Indeed, platforms like StreaMon can take advantage of improved scalability in highly dynamic scenarios, better flexibility and new levels of automation coming from such trend.
 

In order to face these new challenges, in this paper we propose D-StreaMon, a distributed network monitoring framework based on StreaMon \cite{bianchi2014streamon} data-plane programming abstraction for stream-based monitoring tasks directly running over network probes. The building block of this framework is StreaMon, a network monitoring tool based on a data-plane abstraction devised to scalably decouple the ”programming logic” of a traffic analysis application (tracked states, features, anomaly conditions, etc.) from elementary primitives (counting and metering, matching, events generation, etc), efficiently pre-implemented in the probes, and used as common instruction set for supporting the desired logic. Having D-Streamon as main foundation, we realized two NFV-based solutions and we assessed the performances.


The contributions of this paper are:
\begin{itemize}
    \item the extension of the StreaMon monitoring solution, separating the centralized configuration and management by the distributed monitoring probes
    \item the introduction of a distributed architecture for StreaMon based on Publish/Subscribe paradigm
    \item the implementation of an Open Source GUI for the configuration and management system, based on Ansible/Semaphore
    \item the deployment of the Distributed StreaMon (D-StreaMon) over an NFV environment in two different variants
    \item the performance evaluation of the NFV based solutions; 
\end{itemize}

On top of StreaMon framework, the experimenters are able to design and deploy new security features and to experiment on new deployment scenarios with minimal effort. Moreover, it easy to integrate the solution in the NFV framework envisaged by the ETSI group for NFV~\cite{nfv}. The paper is structured as follows: section \ref{sec:streamon} elaborate on StreaMon platform; the D-StreaMon solution is described along with the NFV based solutions in section \ref{sec:dstreamon}; section \ref{sec:experimentalsetup} provides a performance evaluation of the designed solutions; section~\ref{sec:relatedwork} reports on related work; finally in Section~\ref{sec:conclusions} we draw some conclusions and highlight the next steps.

%% file: inc/streamon.tex
\section{StreaMon}
\label{sec:streamon}
StreaMon is a software defined platform for stream-based monitoring tasks directly running over network probes. StreaMon's strategy closely resembles the one pioneered by Openflow \cite{mckeown2008openflow} in the abstraction of networking functionalities, thus paving the road towards software-defined networking. However, the analogy with Openflow limits to the strategic level; in its technical design, StreaMon significantly departs from Openflow for the very simple reason that the data-plane programmability of monitoring tasks exhibits very different requirements with respect to the data-plane programmability of networking functionalities, and thus mandate for different programming abstractions.

This drives towards a {\em different} pragmatic abstraction with respect to a match/action table: \emph{first}, the ``entity'' being monitored is not consistently associated with the same field composition in the packet header (e.g: 2 way communication to/from the same IP address). \emph{Second}, the type of analysis (and possibly the monitoring entity target) entailed by a monitoring application may change over time, dynamically adapting to the knowledge gathered so far. \emph{Third}, activities associated to a monitoring task are not all associated to a matching functionality, but they rather require \emph{triggering conditions} applied to the gathered features. 

\begin{figure*}[t]
\centering
   \includegraphics[width=17cm, height=3cm]{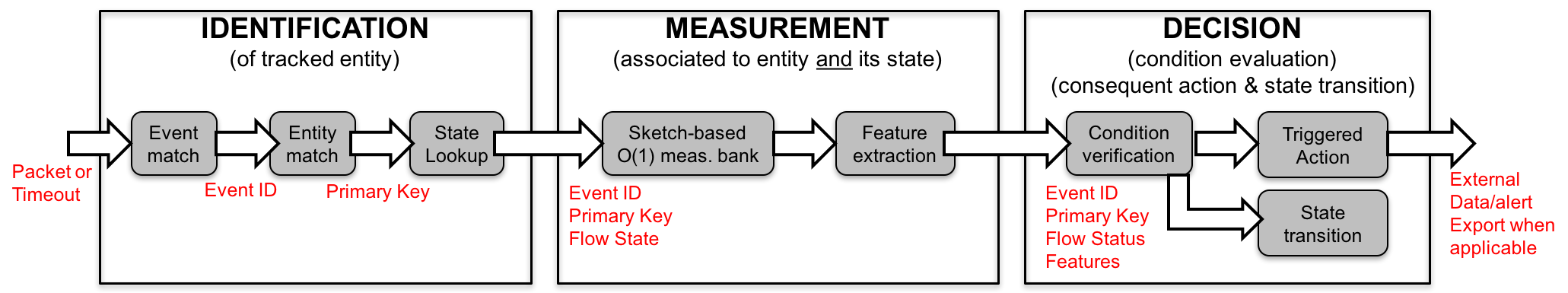}
\caption{StreaMon data plane identification/measurement/decision abstraction, and its mapping to implementation-specific workflow tasks}
\vspace*{-.5cm}
\label{fig:streamon-abstraction}
\end{figure*}

StreaMon's abstraction, illustrated in Figure \ref{fig:streamon-abstraction}, appears capable to cope with such requirements. It consists of three ``stages'', programmable by the monitoring application developer via external means (i.e. not accessing the internal probe platform implementation).

{\bf (1)} The {\bf Identification} stage permits the programmer to specify what is the monitored entity (more precisely, deriving its {\bf primary key}, i.e., a combination of packet fields that identify the entity) associated to an event triggered by the actual packet under arrival, as well as retrieve an eventually associated state.

{\bf (2)} The {\bf Measurement} stage permits the programmer to configure which information should be accounted. It integrates {\em efficiently implemented} built-in hash-based measurement primitives (metric modules), fed by configurable packet fields, with externally programmed {\em features}, expressed as arbitrary arithmetic operations on the metric modules' output. 

{\bf (3)} The {\bf Decision} stage is the most novel aspect of StreaMon. It permits to define the application logic in the form of eXtended Finite State Machines (XFSM), i.e. check conditions associated to the current state and tested over the currently computed features, and trigger associated actions and/or state transitions. 

The previously introduced abstraction can be concretely implemented by a stream processing engine consisting of four modular layers descriptively organized into two subsystems: the {\em Measurement subsystem} and the {\em Logic subsystem}. 

\textbf{Event layer} - Such layer is in charge of parsing each raw captured packet and match an {\em event} among those {\emph user-programmed} via the StreaMon API. The matched event identifies a user-programmed \textit{primary key} which permits to retrieve an {\em eventually} stored state. The event layer is further in charge of supplementary technical tasks, such as handling special timeout events, deriving further secondary keys, etc.

\textbf{Metric layer} - StreaMon operates on a per-packet basis and does {\em not} store any (raw) traffic in a local database. The application programmer can instantiate a number of {\em metrics} derived by a basic common structure, implemented as computation/memory efficient multi-hash data structures (i.e., Bloom-type sketches, d-left hash tables), updated at every packet arrival. 

\textbf{Feature layer} - this layer permits to compute user-defined arithmetic functions over (one or more) metric outputs. Whereas metrics carry out the bulky task of accounting {\em basic} statistics in a scalable and computation/memory efficient manner, the features compute {\em derived} statistics tailored to the specific application needs, at no (noticeable) extra computational/memory cost.

\textbf{Decision layer} - this final processing stage implements the actual application logic. This layer keeps a list of \textit{conditions} expressed as mathematical/logical functions of the feature vector provided by the previous layer and any other possible secondary status. Each condition will trigger a set of specified and built-in \textit{actions} and a state \textit{transition}.

Application programmers describe their desired monitoring operations through an high-level XML-like language, which permits to specify custom (dynamic) states, configure measurement metrics, formalize when (e.g.in which state and for which event) and how (i.e. by performing which operations over available metrics and state information) to extract features, and under which conditions trigger relevant actions (e.g. send an alert or data to a central controller). We remark that a monitoring application formally specified using our XML description does not require to be {\em compiled} by application developers, but is run-time installed, thus significantly simplifying on-field deployment. 


%% file: inc/dstreamon.tex
\section{D-StreaMon}
\label{sec:dstreamon}

The first implementation of StreaMon platform presented a number of limitations which do not make it widely applicable. In this section we elaborate on the evolution of the platform the so called Distributed StreaMon. It introduces a number of novelties which make the platform suitable for use cases like NFV framework and also for new generation threats.

\subsection{Requirements}
\label{sec:requirements}

The StreaMon implementation relies on slow deployment process, which should be streamlined. It should be fast and light in order to react quickly to the network threats. It should be suitable for use cases, like NFV, which request the support for highly dynamic scenarios, in which StreaMon probes should be instantiated ``on the fly'' following the users requests and the networ behavior. Industrial control systems or critical infrastructures requires for a new generation of security monitoring framework like the one envisaged by the SCISSOR project \cite{scissor}. In these challenging scenarios, there is the need for programmable abstractions in the monitoring framework in order to have programmable analysis functions. A control and coordination layer able to adaptively orchestrating remote probes is a mandatory requirements. Another question that should definitely addressed for the next generation monitoring framework is ``How to support resources constrained devices and computationally limited VMs ?''. StreaMon and in general probes should be able to run on low performance devices like network switches or in light-weight virtual computing resources like Unikernels, Tinified VMs and Containers. This will enable the probes to be runnable also in small ubiquitous clouds. The legacy StreaMon considered an implementation through high performance middleboxes which monitor a single-point of the network. Network threats tends to be increasingly complex and distributed. The interconnection and the cooperation of multiple network probes that control different points of the network can be of great value. External components, like the Control and Coordination Layer proposed by the SCISSOR project, need the availability of the monitoring information generated by StreaMon platform. On one hand, the interconnection, through chaining, of multiple StreaMon probes for different levels of processing can improve the whole monitoring performances. On the other hand it reduces the resource requirements making also affordable the deployment in resources limited computing nodes. In this way, probes tend to become small and highly specialized encouraging also the re-usability. In order to benefit of the inherent flexibility and better level of automation introduced by NFV architecture, the management (deployment, configuration, etc.) of multiple StreaMon probes should be programmable and adaptive to the network behavior.

\subsection{Architecture}
\label{sec:architecture}

Legacy StreaMon foresees for a single host architecture, where the single node executes all the steps of the platform life-cycle. It receives in input the user defined xml file which describes the features requested for the monitoring. Then, it parses this file and generates the low-level configuration file for the probe. In parallel, a dynamic library with all the requested features is created. Finally, it runs the probe giving input the two outputs of the previous phases. With D-StreaMon, we introduce a separation of concerns. The new architecture envisages for a Master machine and a number of Slaves machines. In the Master, we move all the static steps like generation of libraries and configuration files while we allocate only the operative steps to the Slave nodes. We rely on cross-compilation to generate probes and then run on the Slave nodes what is generated by the Master. The D-StreaMon archtecture decouples the parsing and compile operations from the execution of StreaMon probes. The Master machine needs the Python interpreter and the g++ compiler; the Slaves need only the libraries necessary to run the StreaMon probes. Thanks to this approach, the master machine is equal to the \textit{full} legacy StreaMon machine. Instead, the Slave machines are very light-weight. The architecture does not exclude multiple Master but at time of writing we are not implemented any replication solution for the Master. 

\begin{figure}
    \centering
    \includegraphics[width=0.40\textwidth]{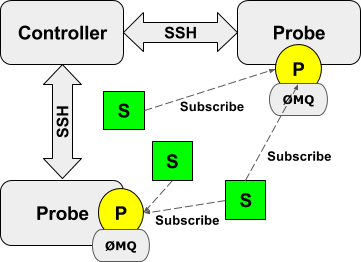}
    \caption{D-StreaMon architecture}
    \vspace{-3ex}
    \label{fig:1}
\end{figure}

As regards the model of the architecture we did not rely on a simple Client and Server scheme. The probes need to communicate and cooperate, the data has to be made available for different types of consumers. The probes can be intermittent and not always connected to the whole system. For these reason, we designed an architecture based on the Publish/Subscribe paradigm. Each probe is an independent Publisher and, according to the configuration file, publishes only some types of events; Subscribers listen only for events of interest. This architecture has been realized using the ZeroMQ \cite{zeromq} library. It is important to note that a Publisher machine can in turn subscribe for events generated by other Publisher machines. The Master is also in charge to monitor all the Slaves nodes. D-StreaMon uses Ansible \cite{ansible} as configuration and monitoring platform. It manages the nodes using the SSH protocol or via PowerShell. The new platform is shipped also with a GUI that enables the installation, removal and monitoring probes in the SLAVE machines. The GUI is based on Semaphore, an open source project that is an alternative to the official GUI Ansible (Tower). It uses Node.js as well as HTML and CSS.

\subsection{NFV and SDN based deployment}
\label{sec:nfvbased}
\begin{figure}
    \centering
    \includegraphics[width=0.48\textwidth]{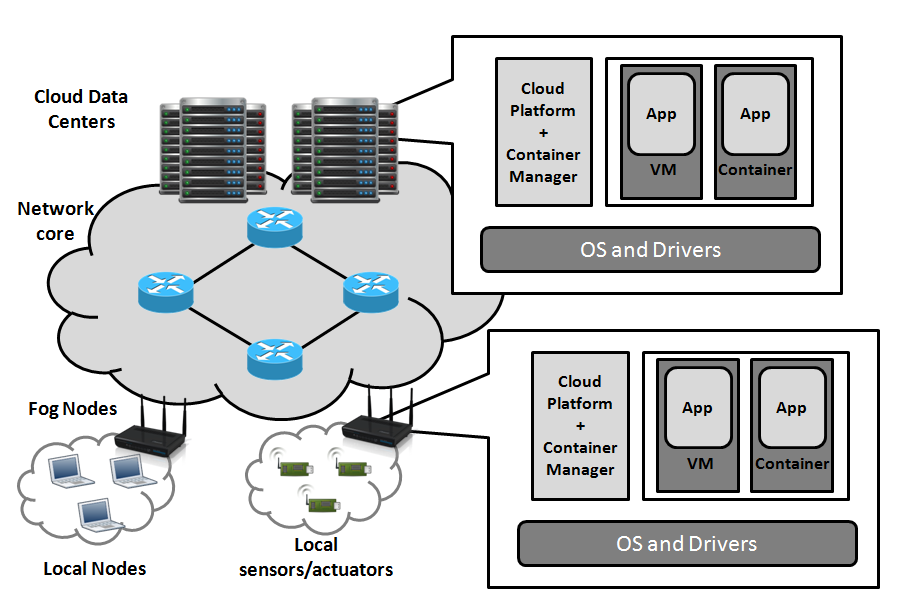}
    \caption{Reference scenarios}
    \vspace{-3ex}
    \label{fig:scenario}
\end{figure}

According to the NFV paradigm, a given service can be decomposed into a set of software components which are executed in virtual computing resources themselves running in virtualization servers. The latter can be deployed in data-center which are located in different sections of the Service Provider's network or in the infrastructure of a Cloud Service Provider. In this paper, we focus on StreaMon applicability in two reference scenarios: Mobile Edge Computing/Fog Computing and Cloud Computing. Figure \ref{fig:scenario} shows the reference scenarios. In the bottom part of the figure, a wireless use case is shown where the base stations offer connectivity to the mobile clients and have also the computing resources to build small scale data center as envisaged by Cloudlets architecture \cite{satyanarayanan2009case}. The applicability of this use case finds solid foundation in recent solutions like NuvlaBox \cite{nuvlabox} which provides a small cloud packaged in a box and is able to support from Fog Computing use case to Internet of Things use case. In the top part of the figure, it is represented the classical Cloud Computing scenario, where the providers guarantee IAAS solution to their users. In this scenario, it is of interest the monitoring of traffic among the virtual computing resources. In this paper, we consider a virtualization solution based on the so called \textit{Containers}, in particular Dockers \cite{docker}.

\begin{figure}
\centering
\begin{subfigure}{0.24\textwidth}
  \centering
  \includegraphics[width=1.05\textwidth]{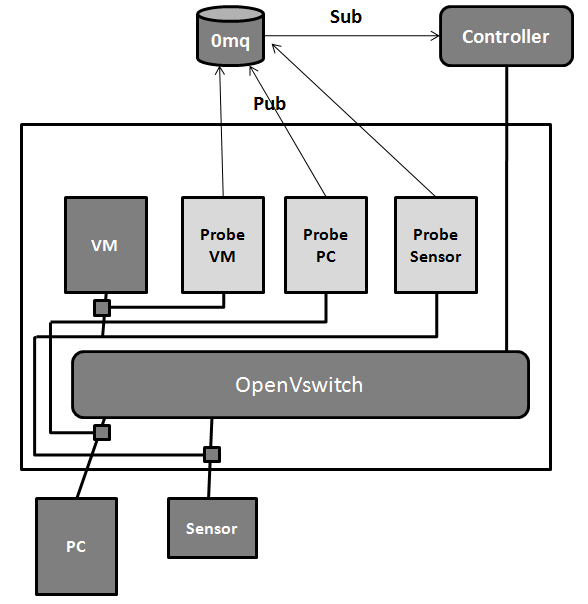}
  \caption{Probes as Processes}
  \label{fig:sub2}
\end{subfigure}
\begin{subfigure}{0.24\textwidth}
  \centering
  \includegraphics[width=1.05\textwidth]{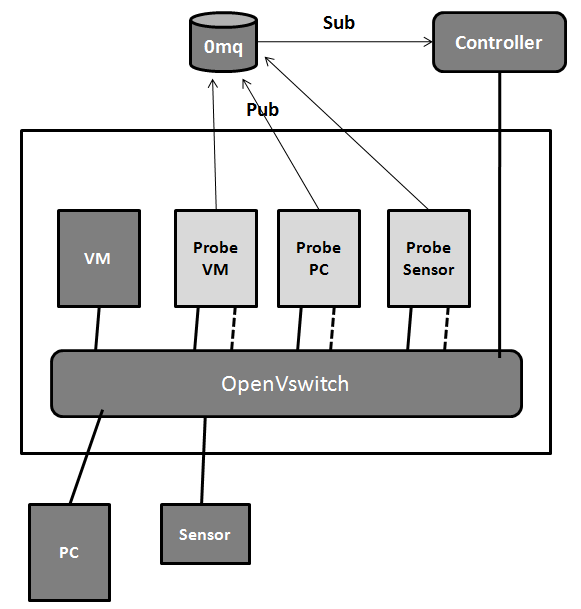}
  \caption{Probes as Containers}
  \label{fig:sub1}
\end{subfigure}
\caption{D-StreaMon architectures}
\label{fig:test}
\end{figure}

Let us consider the migration of the StreaMon platform to a Container based NFV environment. The StreaMon probes needs to be instantiated in Cloud data centers or in small scale cloud at the edge of the network with the objective of monitoring the tenants' Containers. Using the architecture described in section \ref{sec:dstreamon}, we have designed two NFV based deployments. The first solution is referred to as \textit{Probes as Processes} (figure \ref{fig:sub2}). The monitoring probes are executed as processes running in the main host process space. The Probes intercept the traffic directly from the ports of the target Containers which have been bridged in the L2 switch of the virtualization server (in our case we use Open vSwitch \cite{ovs} to implement the Layer 2 switch of the virtualization server). The second solution is referred to as \textit{Probes as Processes} and is represented in figure \ref{fig:sub1}. It envisages the use of the Docker Containers for running the probes: for each Container to be monitored, a StreaMon Probe is created and deployed in a Docker Container. All Containers are attached to the same L2 switch and port mirroring functionality is used to "copy" the VM traffic towards the associated Probe.  The first NFV solution can provide better performance, because the Process virtualization introduces less overhead with respect to Containers solution and because the packets do not need to be "mirrored" by the switch. At the same time the second solution can provide a better isolation and does not request particular workarounds to have different StreaMon Probes working in several Processes co-located in the same host. Moreover, the Container based solution is more attractive for the availability of different orchestration tools like Docker Swarm \cite{docker}, Kubernetes \cite{kubernetes}, Nomad and many others. These solutions can provide valuable improvements to the architecture as they can automate the deployment phase of the Probes and can provide also an interesting solution for the integration in the NFV ecosystem (see \cite{nfv} for an a thorough description of NFV architecture).

%% file: inc/experimentalresults.tex
\section{Experimental Results}
\label{sec:experimentalsetup}

\begin{figure}[!ht]
\centering
    \includegraphics[width=0.5\textwidth]{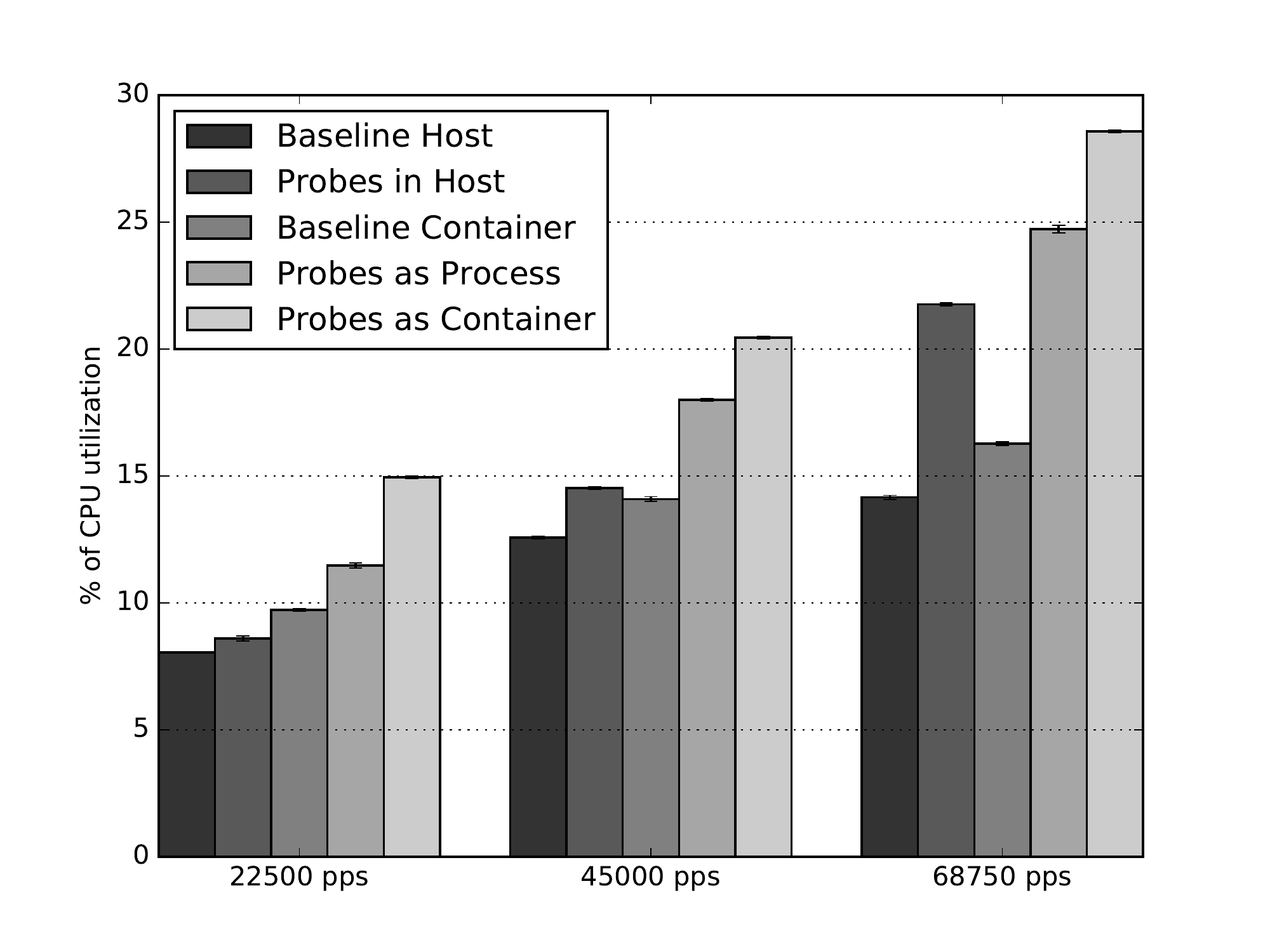}
    \caption{Cpu Usage}
    \label{fig:exper}
\end{figure}

In order to evaluate the suitability of NFV based solutions, we performed several experiments in which we analyzed the CPU load for the different realized solutions. The rationale for this evaluation is to provide an indication on the scalability of the approaches for a Network Function Virtualization Infrastructure made up of Linux VMs (tenants VMs) and Linux Containers/Processes running on typical Virtualization Servers. Our experimental set-up is composed by two hosts with an Intel i3-2100 dual-core CPU and 8G of RAM. One host (hereafter called as HostS) is used as traffic generator, the other host (HostC) as traffic sink and to run D-StreaMon probes. We are using Debian 8.3 operating systems with Xen-enabled v3.16.7 Linux kernels. Both hosts are equipped with two network interfaces at 1 Gb/s, one interface is used as the management interface and the other one for the direct interconnection of the host (data plane network). The CPU load has been obtained gathering the idle percentage through \textit{top} suite and subtracting this value from 100\%. All results have been obtained by executing a test of 100 replicated runs. Error bars in the figures denote the 95\% confidence intervals of the results. However, they are so close to the average that are barely visible in figure \ref{fig:exper}. During the experiment, the probes receive traffic at three different rate of packet/s respectively 22500 pps, 45000 pps and 68750 pps with a packet size of 100 byte. To generate the traffic we used the \textit{iperf} tool. The traffic generator is connected to the remote host and simultaneously launches four different UDP flows. 

For each of the three rates, we have evaluated the CPU utilization in 5 different cases. In the \textit{Baseline host} case, an \textit{iperf} Server is executed directly into HostS and no monitoring probe is active. In the \textit{Probe in host} case we run the StreMon probe in the host, without any virtualization mechanism. In the \textit{Baseline container} the \textit{iperf} Server is executed inside a Docker Container, without no monitorin probes. As we expected, increasing the packets rate introduces a linear increment of the CPU load. Instead, moving the \textit{iperf} server inside a container (\textit{Baseline container}) introduces about 13.17\% of overhead. This is due to the execution of the container itself, the introduction of an Open vSwitch for the forwarding of the traffic and the forwarding of the traffic through the switch. Further analyzing the graph, we can easily argue that the percentage of the overhead introduced is stable while varying the packet rate.

The \textit{Probes as Process} and \textit{Probes as Container} columns shows the CPU utilization of the server machine while running a probe using the NFV based architectures. Obviously, the introduction of the StreaMon probe increases the CPU utilization. The probes have to analyze all the received traffic and, increasing the packets per second, the CPU utilization increases more rapidly than the first two columns. As expected running a probe inside a container is more expensive than running it as a process. The overhead introduced by the container and by the forwarding of the traffic account for about a 15,29\% of overhead. Note that the probes in the containers need two interfaces and the forwarding of the traffic is realized through the mirror functionality provided by Open vSwitch. Also in this case, the overhead is stable when varying the monitored traffic rate.

Analyzing the CPU load results, we can state that running containers in a low-end virtualization add a reasonable low overhead, despite the packet rate increases. The most important result comes out from the comparison of the two NFV based solution: we can run StreaMon probes in Containers with the introduction of a total overhead about 15,29\% respect of process solution. Given the limited amount of the overhead, we believe that the solution we designed are promising solutions for Cloud environment: Service providers can deliver with the same infrastructure (consolidation) both IaaS and the network monitoring without the need for a separated infrastructure of middlebox nodes (different management tools, low level of automation and so on). 

%% file: inc/relatedwork.tex
\section{Related Work}
\label{sec:relatedwork}

Several monitoring platforms have targeted monitoring applications’ programmability. CoralReef \cite{keys2001architecture}, FLAME \cite{anagnostakis2002open} and Blockmon \cite{huici2012blockmon} are frameworks which grant full programmability by permitting the monitoring application developers to “hook” their custom C/C++/Perl traffic analysis function to the platform. Opensketch \cite{yu2013software} proposes an efficient generic data plane based on programmable metric sketches. If on the one hand StreaMon share with Opensketch the same measurement approach, on the other hand its data plane abstraction delegates any decision stage and logic adaptation the control plane and, with reference to our proposed abstraction, does not go beyond the functionalities of our proposed Measurement Subsystem. On the same line, ProgME \cite{yuan2011progme} is a programmable measurement framework which revolves around the extended and more scalable notion of dynamic flowset composition, for which it provides a novel functional language. Even though equivalent dynamic tracking strategies might be deployed over Openflow based monitoring tools, by exploiting multiple tables, metadata and by delegating ”monitoring intelligence” to external controllers, this approach would require to fully develop the specific application logic and to forward all packets to an external controller, (like in Openflow based monitoring tool Fresco \cite{shin2013fresco}), which will increase complexity and affect performance.

Monitoring solutions based on NFV and SDN concepts is a topic addressed also by other works. Authors in \cite{zaalouk2014orchsec} design an orchestrator based solution that leverages on the functionalities of a Network Monitor and SDN controller to execute security applications. Respect to our work, this solution does not consider neither NFV solutions nor the network monitoring inside the network. Virtualized functions for security appliances are described in \cite{basak2010virtualizing}. However this work does not take into consideration SDN for controlling network operation and does not consider tiny virtual resources to run security functions. CloudSec \cite{ibrahim2011cloudsec} share with our work only the purpose of the real-time security monitoring for the hosted VMs in the IaaS cloud platform. From an architectural point of view is completely different, as it is a monitoring appliance based on introspection techniques. \cite{5359505} described a distributed intrusion detection system which spreads its Agents in the Cloud. This agents run in VMs. Also PsycoTrace \cite{baiardi2009transparent}, \cite{payne2008lares} and other works like \cite{chiueh2009stealthy} exploit virtualization to introduce monitoring solution in the cloud. The approach of running security functions in virtual computing resources is similar to our, but they use full-fledged VMs instead of Containers. Moreover, these works does not consider reconfiguration of the network through SDN as reaction to possible threats. Although, different monitoring solution are based on NFV and SDN concepts, to the best of our knowledge none integrate NFV and SDN in an unique platform like D-StreaMon platform does.

%% file: inc/conclusions.tex
\section{Conclusions}
\label{sec:conclusions}

We have described D-StreaMon an NFV-capable distributed framework for network monitoring. D-StreaMon has been realized on top of StreaMon platform, which is a middlebox solution for network monitoring. A distributed architecture based on Publish/Subscribe model is the foundation of the new implementation. The distribution and separation of concerns delivers also a streamlined deployment procedure. Using this architecture we implemented two NFV based solutions: Processes based and Containers based. We have realized a performance evaluation of the realized solutions. As we expected the amount of the overhead introduced by the NFV designs is limited making them are an attractive solution. Cloud providers can deliver with a unique infrastructure (consolidation) both IaaS and the network monitoring without the necessity for the deployment of a separated infrastructure of middlebox nodes (different management tools, low level of automation and so on). This will open also a new interesting scenarios where the network monitoring can be provided to the users as a service (XaaS) thanks also to the programmability functionalities of platforms like StreaMon. All the source code of the developed solution is available at \cite{dstreamonrepo}

Future works will address: i) the improvement of the performances of the D-StreaMon probes in terms of packet processing; ii) the realization of a coordinated solution which integrates NFV, SDN and Cloud Computing in one infrastructure. Improve the performances of the probes is an important step as currently we have a limit on the number of packets that the probes are able to handle. A first step towards the resolution could be the substitution of \textit{Libpcap} library at the base of StreaMon platform with a more performant solution. The realization of an integrated SDN, NFV and Cloud infrastructure will provide the final step towards a fully functional solution. The Cloud OS can guarantee the automated deployment of probes and can provide useful information to the SDN controller about tenants VMs and associated probes. These information are useful for the Network OS to immediately react if a network probe notifies for malicious actions.